\documentclass[preprint,showpacs,preprintnumbers,amsmath,amssymb]{revtex4}

\usepackage{graphicx}
\usepackage{dcolumn}
\usepackage{bm}

\def\l{\left}
\def\r{\right}

\newcommand{\bra}[1]{\ensuremath{\langle #1 |}}
\newcommand{\ket}[1]{\ensuremath{| #1 \rangle}}
\newcommand{\half}{\ensuremath{\frac{1}{2}}}
\newcommand{\expect}[1]{\ensuremath{\l\langle #1 \r\rangle}}
\newcommand{\psid}{\ensuremath{\psi^{\dagger}}}

\newcommand{\cycfreq}{\ensuremath{\omega_c}}
\newcommand{\ml}{\ensuremath{\ell_B}}
\newcommand{\cycfreqb}{\ensuremath{\overline{\omega}_c}}
\newcommand{\mlb}{\ensuremath{\overline{\ell}_B}}
\newcommand{\Bb}{\ensuremath{\overline{B}}}

\begin{document}

\title{Effects of interaction induced second Landau level mixing in the $\nu=1$ quantum Hall effect}

\author{Mats Horsdal}
\affiliation{Department of Physics, University of Oslo, P.O. Box 1048 Blindern, 0316 Oslo, Norway}

\date{\today}

\begin{abstract}
Work by Mandal and Jain [S. S. Mandal and J. K. Jain, Solid State Commun. \textbf{118}, 503 (2001)] suggests that interaction induced mixing with the second composite fermion Landau level can lead to renormalization of the electron correlation function exponent in the fractional quantum Hall effect. In the work reported here a similar mixing with the second electronic Landau level is studied in the $\nu=1$ integer case. The ground state is calculated by use of the Hartree-Fock algorithm, and the electron density and electron correlation function on the edge are calculated. It is shown that the interaction gives rise to oscillations in the density profile. In particular, a short range interaction gives a profile qualitatively similar to the results reported by Mandal and Jain. On the other hand, no renormalization of the correlation function exponent is found.
\end{abstract}

\maketitle


\section*{Introduction}

In the past few years there has been renewed interest in the edge physics of the QH system initiated by the discrepancy between the observed tunneling resistance \cite{ChangEtAlPRL12,HilkeAtALPRL87} and a simple application of Wen's chiral Luttinger liquid description of the edge \cite{Wen1990PRB,Wen1992IJMPB}.
According to this description the Luttinger parameters will have universal values determined by the topological properties of the bulk. For a Luttinger liquid the electron correlation function, $|\expect{\psi^\dag(x)\psi(x')}|$, exhibits a power law behaviour, $|x-x'|^{-\gamma}$, in the large $|x-x'|$ limit, where $\gamma$ is determined by the Luttinger parameters. In the simplest case, $\nu=1/(2m+1)$, the correlation function exponent will have a universal value determined by the filling fraction; in the $\nu=1/3$ case it can be shown that $\gamma=3$.
The observed tunneling resistance does indeed confirm a Luttinger liquid description of the edge in the sense that the asymptotic behaviour of the edge correlation function exhibits a power law, but the exponent deviates from the universal value predicted by the general theory.

Numerical works by Mandal and Jain (MJ) \cite{MandalJain2001SSC,MandalJain2002PRL} and Tsiper and Goldman \cite{GoldmanTsiper2001PRL,TsiperGoldman2001PRB} have attributed the discrepancy to the effect of the electron-electron interaction. Their results suggest that the the exponent is not universal, but is renormalized by the interaction.
However, a different view is taken by Wan, Evers and Rezayi \cite{WanEversRezayi2005PRL}, who  suggest that the discrepancy between experiment and theory can be attributed to the details of the edge confinement. In their study, they also found a change of the correlation function due to the interaction, but only in the case of a so-called 'hard edge', which was used by MJ and Goldman and Tsiper. When a 'soft edge' was introduced, or a neutralizing background potential was introduced in the `hard edge` case, the universal exponent was retained. They suggested that the exponent is indeed universal, and that the edge confinement is responsible for the experimental results.

Mandal and Jain \cite{MandalJain2001SSC} studied a system in the circular gauge and introduced an interaction between composite fermions (CF) that induced mixing with the second CF Landau level. In the CF description of the fractional QH effect, the ground state of the system is made up of CFs fully occupying one or more CF Landau levels; i.e. the fractional QH effect is an integer QH effect of CFs. The electronic ground state is then found by multiplying the CF ground state with a Jastrow factor and then projecting onto the lowest Landau level (LLL). 
MJ introduced a Coulomb interaction and found the ground state by exact diagonalization in a Hilbert space spanned by a Laughlin state and states with a single CF excited to the corresponding angular momentum state in the second CF Landau level. This corresponds to a 'hard edge' since scattering to higher angular momentum states is not allowed.
They found that the interaction gave rise to oscillations in the density profile and to renormalization of the asymptotic correlation function exponent. A Yukawa-type interaction also renormalized the exponent, but with a value different from the Coulomb case. They suggested that the exponent is not universal and its value is determined by the details of the interaction.

We see no obvious reason why a similar renormalization effect should not be present in the $\nu=1$ case if the electron-electron interaction causes mixing with the second electronic Landau level. To see whether such an effect is present, we study a model with linear geometry. Each single particle orbital in a completely filled LLL is allowed to mix with the same momentum orbital in the second Landau level (2LL). Scattering to higher momentum states is not allowed, which corresponds to a 'hard edge' condition, as in ref. \cite{MandalJain2001SSC}.
To determine the the ground state wavefunction of the interacting system, we use the Hartree-Fock (HF) algorithm, rather than exact diagonalization.
MJ's results suggests that the interaction form is not important for a renormalization to take place. 
For that reason a Gaussian interaction is chosen here, a choice which gives an easy control of both the interaction length and interaction strength.

An advantage of working in the $\nu=1$ regime is that a calculation involving a large number of particles can easily be done. This, combined with the linear gauge, means that large system lengths can be achieved since the system length is proportional to the number of particles and $k_F^{-1}$.
The calculation reported here involves 1001 particles and the Fermi momentum is $k_F=10\ml^{-1}$. This gives rise to a system length of about 300 magnetic lengths.
The choice of linear geometry gives rise to two edges. Effects of the edge-edge interaction in the $\nu=1$ regime will not be discussed here since it has been studied previously \cite{HorsdalLeinaas02PRB2007}. The cited work only included states in the LLL and scattering to higher momentum states were allowed, which makes the work complementary to the study reported here, where only mixing with the same momentum state in the 2LL is allowed.
To mimic the positive background potential found in real systems a harmonic potential is introduced, and it is used to compensate for the expansion of the system due to the repulsive particle interaction.

The electron density and the electron correlation function are calculated in the ground state. We find that that the interaction gives rise to oscillations in the density profile. For a short range interaction the density, close to the edge, qualitatively resembles the edge profile reported in  \cite{MandalJain2001SSC} and \cite{TsiperGoldman2001PRB}. For the correlation function no renormalization of the asymptotic exponent is found.


\section*{Model}
As already discussed we will work in the linear gauge where the vector potential is given by $A_x=-yB$, $A_y=0$. The single particle Hamiltonian can then be written as
\begin{eqnarray}
h&=&\frac{1}{2m}(\mathbf p -e\mathbf A)^2 \nonumber \\
&=&\frac{1}{2m} \l(p_y^2+m\cycfreq^2\l(y+\frac{\ml^2}{\hbar}p_x\r)^2\r),
\label{eq:ham}
\end{eqnarray}
where it is assumed that $eB>0$ and the magnetic length, $\ml=\sqrt{\hbar/eB}$, and the cyclotron frequency, $\cycfreq=eB/m$, have been introduced.
The Hamiltonian, (\ref{eq:ham}), is explicitly translationary invariant in the x-direction and energy eigenstates that are also eigenstates of $p_x$ can therefore be defined. With the momentum quantized as as $p_x=\hbar k$, the Hamiltonian is reduced to a one-dimensional harmonic oscillator with potential minimum at the $k$-dependent position $y_k=-\ml^2k$. The energy eigenstates can be written as
\[
\psi_{kn}(x,y)= {\cal N} e^{ikx} \psi_n(y-y_k),
\]
where $\cal N$ is an appropriate normalization factor and $\psi_n$ is the $n$'th energy eigenstate of a harmonic oscillator. Note that the system is degenerate with respect to $k$.

Let us now introduce a harmonic background potential in the $y$-direction:
\begin{eqnarray}
h&=&\frac{1}{2m}(\mathbf p -e\mathbf A)^2 +\half m\cycfreq^2y^2 \nonumber \\
&=&\frac{1}{2m}p_y^2 
+ \half m\cycfreqb^2\l(y+\frac{\cycfreq}{m\cycfreqb^2}p_x\r)^2
+\frac{1}{2m}\frac{\cycfreq^2}{\cycfreqb^2}p_x^2.
\label{eq:hamBack}
\end{eqnarray}
In the last line an effective cyclotron frequency, $\cycfreqb=e\Bb/m=\sqrt{\omega^2+\cycfreq^2}$, with an effective magnetic field $\Bb=\sqrt{B^2+m\omega^2/e^2}$, has been introduced. We see that the first two terms in (\ref{eq:hamBack}) can be identified as the Hamiltonian of an electron in the effective magnetic field $\Bb$. Since the $x$-momentum is still a constant of motion it can be quantized, and the energy eigenstates will have the same form as before, except for a modified magnetic length $\mlb=\sqrt{\hbar/e\Bb}$ and with $y_k=-\frac{\omega}{\cycfreqb}\mlb^2k$. The last term in (\ref{eq:hamBack}) lifts the degeneracy in $k$ and the energy eigenvalues are given by
\[
\varepsilon_{kn}=\Lambda\hbar\cycfreq\l(n+\half\r) + \Lambda^{-2}
\l(\frac{\omega}{\cycfreq}\r)^2 \frac{\hbar^2k^2}{2m},
\]
where a dimensionless variable has been introduced
\[
\Lambda=(\ml/\mlb)^2=\cycfreqb/\cycfreq=\sqrt{1+(\omega/\cycfreq)^2}.
\]

If we assume periodic boundary conditions of period $L$ in the $x$-direction, the energy eigenstates corresponding to the LLL and the 2LL can be written as
\begin{eqnarray}
\psi_{kn}(x,y)=&&\l(\frac{\Lambda}{\pi}\r)^{1/4} \sqrt{\frac{1}{\ml L}}
e^{ikx} \l\{\frac{\sqrt{2\Lambda}}{\ml}
\l(y+\frac{\ml^2k}{\Lambda^2}\r)\r\}^n \nonumber \\
&&\times\exp\l(-\l(y+\frac{\ml^2k}{\Lambda^2}\r)^2
\frac{\Lambda}{2\ml^2}\r),
\label{eq:psikn}
\end{eqnarray}
where $n=0,1$. The wavefunction (\ref{eq:psikn}) is centered at $y=-\ml^2k/\Lambda^2$, which means that we have a correspondence  between the $y$-position of the particle and the particle momentum $\hbar k$.

Let us introduce the electron interaction with a Gaussian form, $V(\mathbf r)=V_0e^{-\alpha^2\mathbf r^2}$, which allows for easy control of both the interaction strength, $V_0$, and the interaction length, $1/\alpha$.

We will assume the ground state to have the form;
\begin{equation}
\ket{\mbox{GS}}=\prod_{k=-k_F}^{k_F}
\l( C_0(k)c_{k0}^\dag + C_1(k)c_{k1}^\dag \r) \ket{0},
\label{eq:GS}
\end{equation}
where $c_{kn}^\dag$ creates a particle of momentum $\hbar k$ in the $n$'th Landau level and $C_0(k)$ and $C_1(k)$ are mixing coefficients. Normalisation of the ground state requires that $|C_0(k)|^2+|C_1(k)|^2=1$. This form easily conserves total linear momentum because each single particle orbital only includes states with the same $k$-number.
In the weak interaction limit the mixing with the 2LL will be small and we can assume that $C_0(k)\approx 1$ and $\frac{C_1(k)}{C_0(k)}\ll 1$. In an expansion to first order in $\frac{C_1(k)}{C_0(k)}$ we see that the ground state is a superposition of a filled LLL ground state and states with one particle promoted to the 2LL.
The restriction that all momenta has to be less than $k_F$, corresponds to a 'hard edge'.

The single particle orbitals will be of the form $\varphi_k(\mathbf r)=\sum_{n=0,1}C_n(k)\psi_{kn}(\mathbf r)$. From symmetry about the $x$-axis and the correspondence between $k$ and $y$ we would expect the orbitals to behave as $|\varphi_k(x,y)|=|\varphi_{-k}(x,-y)|$. We see from (\ref{eq:psikn}) that $\psi_{k0}(x,y)$ is symmetric and $\psi_{k1}(x,y)$ is antisymmetric (except for the phase factor $e^{ikx}$) when $(k,y)\rightarrow(-k,-y)$. We would therefore expect $C_0(k)$ to be symmetric and $C_1(k)$ to be anti-symmetric when $k\rightarrow -k$. 

In terms of the ground state (\ref{eq:GS}) the total energy of the system is given by
\begin{eqnarray}
E_{GS}&=&\bra{\mbox{GS}} H \ket{\mbox{GS}} \nonumber \\
&=&\sum_{i=-k_F}^{k_F} \bra{\varphi_i}h\ket{\varphi_i}
+\half \sum_{i,j=-k_F}^{k_F} \l\{
\bra{\varphi_i\varphi_j}V \ket{\varphi_i\varphi_j}
-\bra{\varphi_i\varphi_j}V \ket{\varphi_j\varphi_i}\r\},
\label{eq:EGS}
\end{eqnarray}
where
\[
\bra{\varphi_i}h\ket{\varphi_j}
=\int d\mathbf r \varphi_i^*(\mathbf r)h\varphi_j(\mathbf r)
\]
and
\[
\bra{\varphi_i\varphi_j}V \ket{\varphi_k\varphi_l}
=\int d\mathbf r d\mathbf r' \varphi_i^*(\mathbf r)\varphi_j^*(\mathbf r')
V(\mathbf r - \mathbf r') \varphi_k(\mathbf r) \varphi_l(\mathbf r')
\]
Minimizing the energy (\ref{eq:EGS}) leads to the Hartree-Fock (HF) equation which determines the single particle orbitals
\begin{equation}
\l\{h+\sum_{q=-k_F}^{k_F}(I_q-K_q)\r\}\varphi_k(\mathbf r)
=\lambda(k)\varphi_k(\mathbf r),
\end{equation}
where $h$ is the single particle Hamiltonian and $\lambda(k)$ is an eigenvalue. The operator $I_k$ is defined as 
\begin{equation}
I_k=\int d\mathbf r' \varphi^*_k(\mathbf r')V(\mathbf r - \mathbf r')
\varphi_k(\mathbf r'),
\end{equation}
and the exchange operator $K_k$ by
\begin{equation}
K_k\psi(\mathbf r) = \int d\mathbf r' \varphi^*_k(\mathbf r')
V(\mathbf r - \mathbf r') \psi(\mathbf r') \varphi_k(\mathbf r).
\end{equation}
By introducing $\varphi_k(\mathbf r)=\sum_{n=0,1}C_n(k)\psi_{kn}(\mathbf r)$ into the the HF equation and multiplying with $\int d\mathbf r \psi_{kn}^*(\mathbf r)$ and using the fact that the $\psi_{kn}$'s are orthonormal, the HF equation takes the form
\begin{equation}
\l(\begin{matrix}
F_{00}(k) & F_{01}(k) \\
F^*_{01}(k) & F_{11}(k)
\end{matrix}\r)
\l(\begin{matrix}
C_0(k) \\ C_1(k)
\end{matrix}\r)
= \lambda(k)
\l(\begin{matrix}
C_0(k) \\ C_1(k)
\end{matrix}\r),
\label{eq:HFmtrx}
\end{equation}
where
\begin{eqnarray}
F_{nm}(k)=\delta_{nm}\varepsilon_{nm} &+&\sum_{q=-k_F}^{k_F} \sum_{n_1,n_2=0,1}
C^*_{n_1}(q)C_{n_2}(q) \nonumber \\
&\quad& \times\l\{\bra{kn;qn_1}V\ket{km;qn_2}-\bra{kn;qn_1}V\ket{qn_2,km}\r\}
\label{eq:Fnm}
\end{eqnarray}
It is easily shown that $F_{nm}(k)^*=F_{mn}(k)$, which has been used in the derivation of (\ref{eq:HFmtrx}). The matrix element in (\ref{eq:Fnm}) is defined as
\begin{eqnarray}
&&\bra{k_1n_1;k_2n_2}V\ket{k_1'n_1';k_2'n_2'} \nonumber \\
&&= \int d\mathbf r d\mathbf r' \psi^*_{k_1n_1}(\mathbf r)
\psi^*_{k_2n_2}(\mathbf r')V(\mathbf r -\mathbf r') 
\psi_{k_1'n_1'}(\mathbf r)\psi_{k_2'n_2'}(\mathbf r').
\label{eq:MtrxElmnt}
\end{eqnarray}
In the limit $L\gg \ml$  the $x$-integrals can be performed. By introducing the gaussian interaction the matrix element takes the form
\begin{eqnarray*}
&&\bra{k_1n_1;k_2n_2}V\ket{k_1'n_1';k_2'n_2'} \\
&&= \frac{\Lambda V_0}{\sqrt{\pi}\alpha L} 
(2\Lambda)^{(n_1+n_2+n_1'+n_2')/2} \delta_{k_1+k_2,k_1'+k_2'}
e^{-(k_1-k_2)^2/4\alpha^2} \\
&&\quad\times \int_{-\infty}^\infty dydy' e^{-\alpha^2\ml^2(y-y')^2} \\
&&\qquad\qquad\times (y+\ml k_1/\Lambda^2)^{n_1}
e^{-(y+\ml k_1/\Lambda^2)^2\Lambda/2} \\
&&\qquad\qquad\times (y'+\ml k_2/\Lambda^2)^{n_2}
e^{-(y'+\ml k_2/\Lambda^2)^2\Lambda/2} \\
&&\qquad\qquad\times (y+\ml k_1'/\Lambda^2)^{n_1'}
e^{-(y+\ml k_1'/\Lambda^2)^2\Lambda/2} \\
&&\qquad\qquad\times (y'+\ml k_2'/\Lambda^2)^{n_2'}
e^{-(y'+\ml k_2'/\Lambda^2)^2\Lambda/2},
\end{eqnarray*}
where the integration variables $y$ and $y'$ are dimensionless.

The problem of finding $\mathbf C (k)=(C_0(k),C_1(k))$ has now been reduced to a $2\times 2$ eigenvalue problem by equation (\ref{eq:HFmtrx}).
The problem has to be solved iteratively: One starts by calculating the matrix $\mathbb F(k)$ from the non-interacting ground state, $\mathbf C(k)=(1,0)$. Eq. (\ref{eq:HFmtrx}) is solved for $\mathbf C(k)$ and $\overline{\mathbf C}(k)$ corresponding to the lowest eigenvalue, $\lambda(k)$, and highest eigenvalue, $\overline{\lambda}(k)$, respectively. The new ground state is then constructed from the set of all $\mathbf C(k)$ and the procedure is repeated until convergence of $\mathbf C(k)$ is achieved.
For the final set of all $\mathbf C(k)$ to constitute a ground state of the system, the corresponding energy $E_{GS}$ has to be lower than the energy $E_{q_1,q_2}$ of an excited state constructed by replacing one state $\mathbf C(q_2)$ with $\overline{\mathbf C}(q_1)$. The requirement is therefore that $\Gamma_{q_1,q_2}\equiv E_{GS}-E_{q_1,q_2}>0$ for all $q_1$ and $q_2$, where $\Gamma_{q_1,q_2}$ can be written as
\begin{eqnarray*}
\Gamma_{q_1,q_2}=&&\overline{\lambda}(q_1)-\lambda(q_2)
-\sum_{\{n_i\}=1}^2
\overline{C}_{n_1}^*(q_1)C_{n_2}^*(q_2)\overline{C}_{n_3}(q_2)C_{n_4}(q_2) \\
&&\times\l\{ \bra{q_1n_1;q_2n_2}V\ket{q_1n_3;q_2n_4} 
-\bra{q_1n_1;q_2n_2}V\ket{q_2n_4;q_2n_4} \r\}.
\end{eqnarray*}

Due to the translational invariance in the $x$-direction the electron correlation function will only depend on the relative distance in the $x$-direction, $C(x;y,\eta)=\expect{\psid(x,y+\eta/2)\psi(0,y-\eta/2)}$. In terms of the ground state coefficients $C_n(k)$ and by using the expression for the field operator, $\psi(x,y) = \sum_n \sum_k \psi_{kn}(x,y)c_{kn}$, the corrrelation function is given by
\begin{eqnarray}
C(x;y,\eta)=\sum_{k=-k_F}^{k_F}
&& \l\{C^*_0(k)\psi^*_{k0}(x,y+\eta/2)+C^*_1(k)\psi^*_{k1}(x,y+\eta/2)\r\} 
\nonumber \\
\times &&\l\{C_0(k)\psi_{k0}(0,y-\eta/2)+C_1(k)\psi_{k1}(0,y-\eta/2)\r\}.
\label{eq:corr}
\end{eqnarray}
The electron density is found from (\ref{eq:corr}) by setting $x=\eta=0$;
\begin{equation}
\rho(y)=C(x=0;y,\eta=0).
\label{eq:rho}
\end{equation}


\section*{Results}

For the numerics we have chosen $k_F=10\ml^{-1}$ and $N=1001$ particles, which gives a system length of $L=\pi(N-1)k_F^{-1}\approx 300\ml$. From the correspondence between the position in $y$-space and $k$-space we see that the width of a free system is about $W=2k_F\ml^2=20\ml$. To easily relate the interaction parameters to the system size, we define the interaction length as $\sigma=1/\alpha$, and it is given in units of magnetic lengths. For the rest of this paper all length scales are given in units of the magnetic length and all wave vectors in units of inverse magnetic length. The interaction strength, $V_0$, is given in units of $\hbar\cycfreq$, while the harmonic oscillator frequency is given in units of $\cycfreq$.

The ground states in terms of $C_0(k)$ and $C_1(k)$, the electron density (\ref{eq:rho}) and the electron correlation function at the the edge, $C(x;y=\ml^2k_F,\eta=0)$, have been calculated for three different cases with interaction lengths given by $\sigma=1\ml$, $\sigma=5\ml$ and $\sigma=10\ml$.

To make it easier to compare results with a non-interacting system, it is preferable for the interacting system to have a system width of the same size as the non-interacting case. At the same time the requirement that  $\Gamma_{q_1,q_2}>0$, for all $q_1$ and $q_2$, has to be fulfilled to ensure that the ground state corresponds to a true minimum. To meet these demands both the frequency of the harmonic background potential, $\omega$, and the interaction strength, $V_0$, have been tuned in such a way that both demands are satisfied.

To find the number of HF-iterations required, $C_n(k)$ has been plotted, as a function of $k$, for different numbers of iterations and for a limited number of particles. The lowest number of iterations that does not show a change in $C_n(k)$ from one iteration to the next, is chosen. This number of iterations have then been used for the full HF-calculation with 1001 particles. The number of iterations for the $\sigma=5\ml$ and $\sigma=10\ml$ case is 5, and for the $\sigma=1\ml$ case 9 iterations have been used.

\begin{figure}
\begin{center}
\includegraphics[width=350pt]{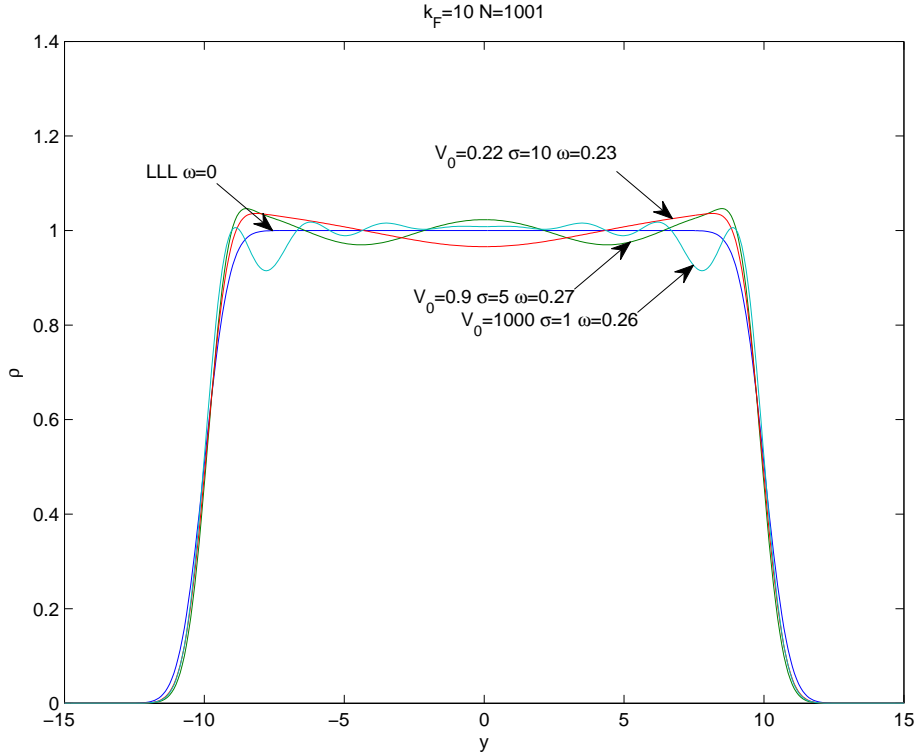}
\caption{The figure shows the density profiles for a system entirely in the LLL with $\omega=0$ and the interacting cases with different interaction lengths. The density is measured in units of the non-interacting bulk density, $1/2\pi\ml^2$.}
\label{fig:Dens}
\end{center}
\end{figure}

Figure \ref{fig:Dens} shows plots of the electron density for a non-interacting system and the three interacting cases with different interaction lengths. We see that the interaction gives rise to oscillations that reach across the whole system, where the oscillation length increases with increasing interaction length. Close to the edge the $\sigma=1\ml$ case qualitatively resembles the edge profile reported in \cite{TsiperGoldman2001PRB} and \cite{MandalJain2001SSC} for the $\nu=1/3$ case. Also the oscillation length, roughly $3\ml$, is of the same magnitude as in the cited references. When the the interaction is increased beyond $\sigma=10\ml$, no qualitative differences can be observed in the density profile. 
A possible explanation  is that the interaction will effectively behave like a constant background potential, when the interaction length is comparable to the system width.

\begin{figure}
\begin{center}
\includegraphics[width=220pt]{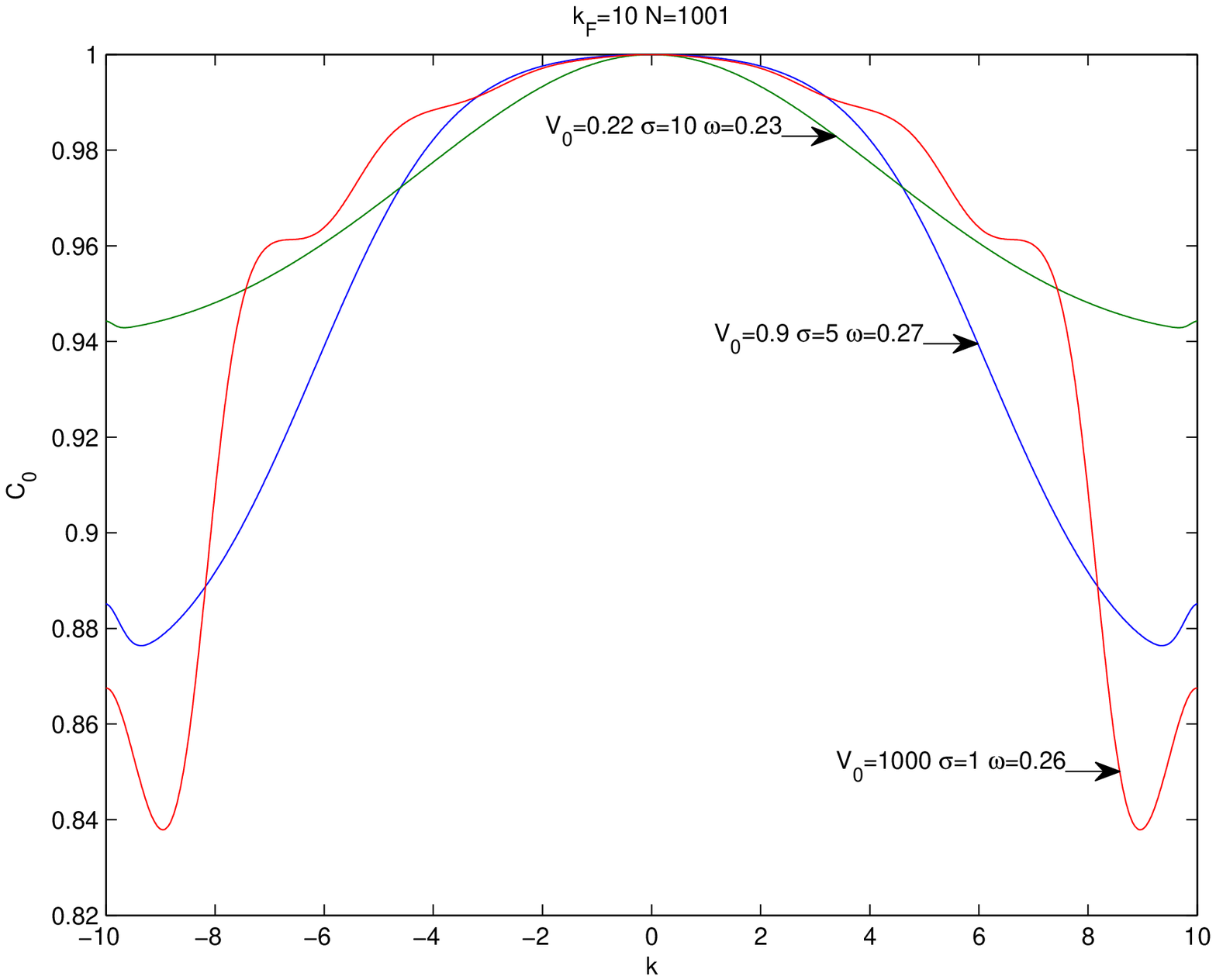}
\includegraphics[width=220pt]{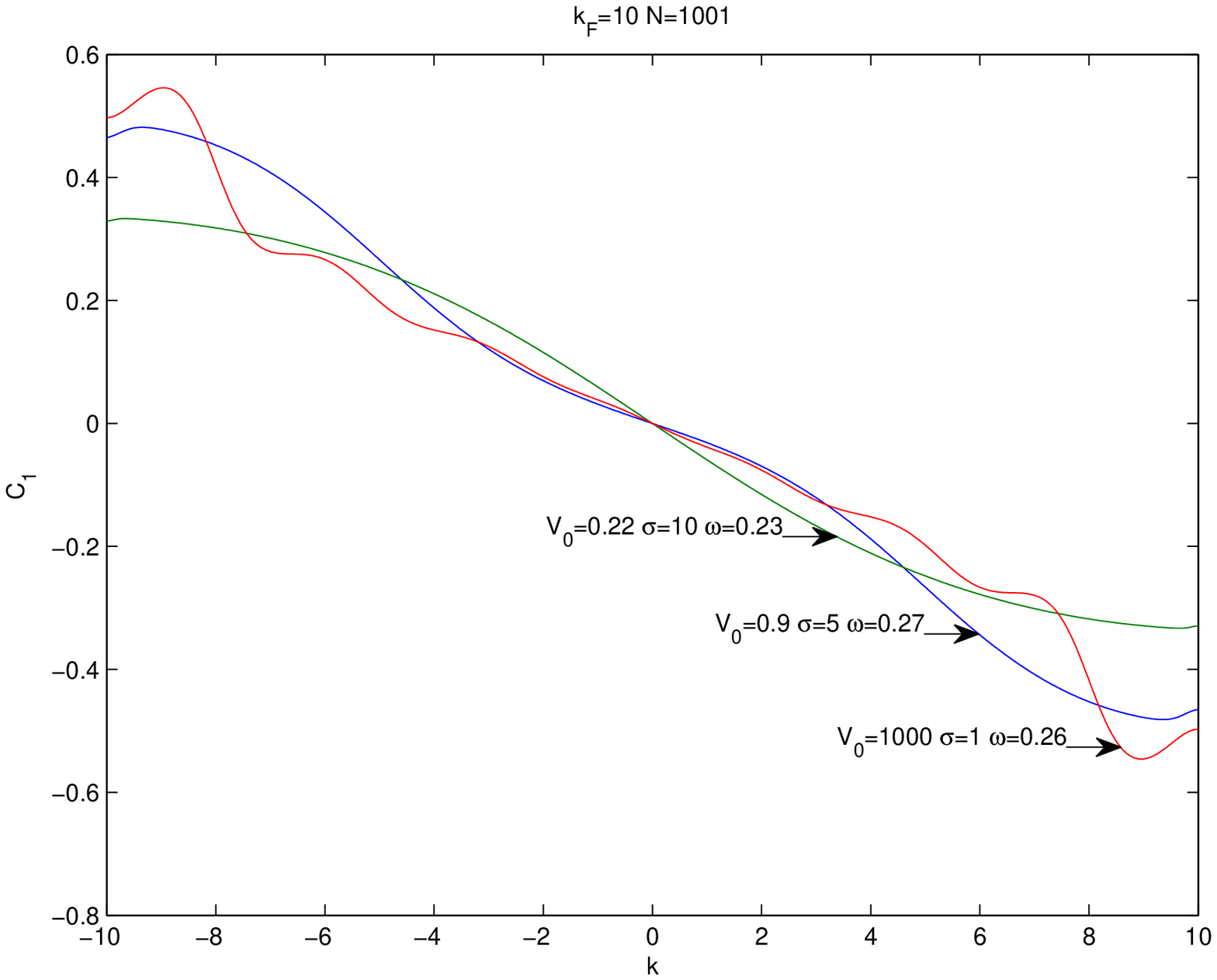}
\caption{Plots of the mixing coefficients $C_0$ and $C_1$ as a function of $k$, in units of $\ml^{-1}$, for the cases shown in Figure \ref{fig:Dens}. The number of HF-iterations is 5 for the $\sigma=5\ml$ and $\sigma=10\ml$ cases and 9 for the $\sigma=1\ml$ case. As pointed out earlier we see that $C_0(k)$ is symmetric about $k=0$ and $C_1(k)$ is antisymmetric.}
\label{fig:C}
\end{center}
\end{figure}

Figure \ref{fig:C} shows plots of the mixing coefficients $C_0$ and $C_1$ as a function of $k$ for the three cases. For the $\sigma=1\ml$ case we see that the oscillations found in the density profile are also present in the mixing between the Landau levels. It is also evident that the relative mixing between the Landau levels is of the same magnitude, even if the variation in $V_0$ between the systems is quite large.

\begin{figure}
\begin{center}
\includegraphics[width=350pt]{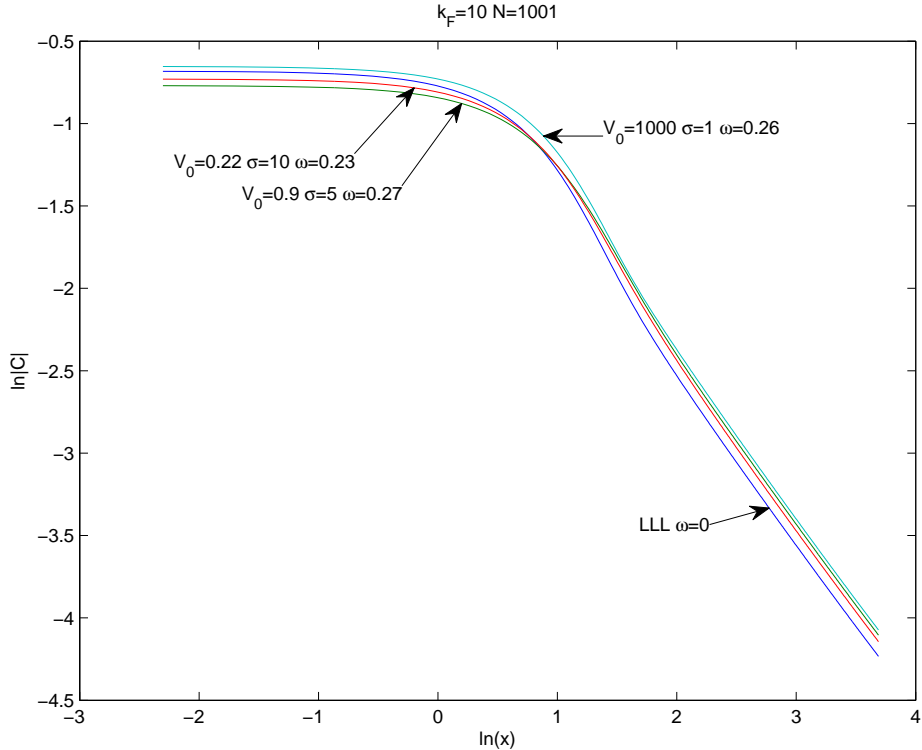}
\caption{The absolute vale of the correlation function at the edge, $|C(x;y=\ml^2k_F,\eta=0)|$, as a function of $x$ for the systems in Figure \ref{fig:Dens}. The correlation function is measured in units of the non-interacting bulk density, $1/2\pi\ml^2$.}
\label{fig:Corr}
\end{center}
\end{figure}

The similarity of our $\sigma=1\ml$ density profile and the results reported in \cite{TsiperGoldman2001PRB} and \cite{MandalJain2001SSC} suggests that we might see a similar renormalization of the exponent of the correlation function at the edge. Fig. \ref{fig:Corr} shows plots of the correlation functions for the different cases, and we see that this is not the case: The three different cases approach the same asymptotic behaviour as the non-interacting case. A linear fit to the log-log plots for a distance from $\sim 8 \ml$ to $\sim 40 \ml$ shows agreement with the asymptotic behavior of the non-interacting system with $\gamma=1$.


\section*{Conclusions}
The study presented here is based on a similar construction as by MJ for $\nu=1/3$, but for $\nu=1$ and with a Gaussian interaction. The ground state, for three different values of the interaction length, was calculated using the HF-agorithm.
The density profile for $\sigma=1\ml$ shows oscillation effects similar to what had been reported by MJ and by Tsiper and Goldman \cite{TsiperGoldman2001PRB}. On the other hand the correlation function exponent shows no sign of a renormalization due to the interaction.
MJ's results suggest that the interaction form is not important for a renormalization effect to take place. In this respect the difference in results between the $\nu=1$ and $\nu=1/3$ case is more likely to be due to fundamental differences between the integer and fractional systems, than to the difference in interaction type. We have no simple explanation for this difference.


\section*{Acknowledgements}
The author wishes to thank Jon Magne Leinaas for helpful discussions and useful comments during this work, and Susanne Viefers for useful comments on the manuscript. This work was supported by the Research Council of Norway.


\end{document}